\documentstyle[12pt,elsart12]{article}
\newcommand{\be}{\begin{equation}}
\newcommand{\ee}{\end{equation}}
\begin{document}
\pagestyle{empty}
\begin{flushright}
ROME prep. INFN n.1134 \\
\end{flushright}
\vskip 2.0cm
\centerline{\bf{INVERSE COMPTON SCATTERING ONTO BBR}} 
\centerline{\bf{IN HIGH ENERGY PHYSICS  AND GAMMA}}
\centerline{\bf{(MeV-TeV)  ASTROPHYSICS}}
\vskip 1cm
\centerline{\bf{ Daniele Fargion $^{1,*}$ and Andrea Salis $^1$ }}
\centerline{}
\centerline{}
\centerline{$^1$ {\it Dipartimento di Fisica,
Universita' di Roma ``La Sapienza", }}
\centerline{$^*$ {\it INFN-Sezione di Roma I}}
\centerline{{\it P.le A. Moro 2, 00185 Rome, Italy. }}
\centerline{}
\centerline{}
\begin{abstract}
We considered the Inverse Compton Scattering (ICS) of charged particles onto 
photons whose distribution is a Black Body Radiation (BBR) deriving the exact 
energy and angular differential distribution in the general case 
and in its most useful expansions. These results can be successfully applied 
in high energy accelerators experiments to evaluate the ICS contribution from 
the thermal photons in the cavity as well as in astrophysics where the ICS of 
cosmic rays plays a relevant role in a variety of phenomena. In particular we 
show how our formulae reproduce the ICS energy spectrum recently 
measured at LEP, how it could be considered a key tool in explaining 
the Gamma Ray Bursts (GRB) , SGRs energy spectrum. Finally we predicted
the presence of a low gamma flux, nearly detectable at hundred of TeV
 from SNRs SN1006 as well as, at lower energy (tens TeV, due to
 gamma ray cascading in cosmic BBR), from relic extragalactic highest 
cosmic rays sources  born by jets in AGN,as blazars 3C279,Mrk421,Mrk 501.
\end{abstract}
\vskip 1.5cm
\begin{flushleft}
ROME prep. INFN n.1134 \\
23 February 1996
\end{flushleft}
%\date{}
\newpage
\pagestyle{plain}
\setcounter{page}{1}
\section*{Introduction}
\label{sec:intro}

The Inverse Compton Scattering (ICS) plays a relevant role in highest energy 
astrophysics (cosmic rays and gamma astronomy) [1-6] and high 
energy physics (LEP I, LEP II, accelerators) [7-9]. 
Indeed from one hand the ICS of high relativistic cosmic rays (either 
electron at GeV or proton and nuclei at much higher energies) onto 
electromagnetic fields (either cosmological Black Body Radiation (BBR) at 
$T\approx 2.73~K$, interstellar lights, radio waves or even stationary 
magnetic fields) is the source of high energy photons (X, gamma rays) which we 
do observe in the Universe as diffuse or point source, on the other hand the 
ICS is often the main process responsible for the slowing down (i.e. for 
the energy losses) of energetic charged particles; indeed ICS is often the main 
cause of cosmic rays lifetime behaviour as well as of their energy spectrum 
depletion at harder regions (i.e. of their detailed spectrum shape and 
evolution) [5-6]. Moreover, ICS of relativistic jets by 
compact objects onto thermal photons by a star companion in a binary system or 
accretion disk might be, as recently proposed [10], the key process able to 
produce "gamma jets" responsible (by their rotation and blazing into 
different directions) of the puzzling Gamma Ray Bursts (GRB). So even if 
this subject seems apparently settled [3-4] we revisit the ICS process in 
order to obtain an analytic and compact formula able to describe the 
differential energy and angular ICS scattered photon number spectrum of 
relativistic charges onto BBR photon spectrum. One of the main feature of our 
final expressions is that we can easily cover the entire range of energy of the 
ICS energy spectrum. These results improve the Montecarlo simulation because 
the latter must consider only a thin portion of 
the energy range in order to inspect in detail the ICS spectrum [9]. In the 
following we describe our approach and we show some remarkable results.

\section*{The ICS onto BBR spectrum}

We follow a standard procedure to get the general ICS spectrum of a 
relativistic 
charge (electron, proton, nuclei in cosmic rays or in accelerators bunches) 
hitting photons whose distribution is a BBR spectrum. We consider first the 
photon target distribution in the Laboratory Frame (LF) where the BBR is 
isotropic and homogeneous, then we transform it to the Electron Frame (EF) 
where the BBR is still homogeneous but highly anisotropic, therefore we 
evaluate in the EF the usual Compton scattering and 
finally we transform back the diffused differential photon number to the LF. 
The starting photon target distribution, in the LF, is the well-known 
isotropic and homogeneous BBR whose number density per unit energy 
$\epsilon_o$ and solid angle $\Omega_o$ is given by the Planck formula
\be
\frac{dn_o}{d\epsilon_o d\Omega_o}=\frac{2}{(ch)^3}\frac{\epsilon_o^2}
{\Big[\exp\Big(\frac{\epsilon_o}{\kappa_B T}\Big)-1\Big]}~~~~~~~~~.
\ee
where $\kappa_B$ is the Boltzmann constant and T the BBR temperature. 
We transform this distribution to the EF by standard Lorentz boosts 
choosing as z axis the direction coincident with the initial electron 
momentum and 
reminding that $dn_o/d\epsilon_o$ is a relativistic invariant [2]. In the 
following we label by a $*$ the quantities related to the electron frame EF, 
by a subscript 0 if they are considered before the scattering, by a subscript 
1 if after; so we have
$$
\cos\theta_o=\frac{\cos\theta_o^*+\beta}{1+\beta\cos\theta_o^*}~,~\varphi_o=
\varphi_o^*~,~\epsilon_o=\gamma\epsilon_o^*(1+\beta\cos\theta_o^*)
$$   
where $\beta$ is the adimensional electron velocity and $\gamma$ the 
corresponding Lorentz factor. The transformed BBR number density distribution 
exhibits, in this relativistic limit and in the EF, a clear dipole anisotropy 
and it is becomes more and more peaked around $\theta_o^*=\pi$ as $\gamma$ 
increases. An analogous dipole signature, at non relativistic regime, is the 
cosmological one found at millikelvin level in the $T=2.73~K$ BBR due to the 
Earth motion. The associated energy spectrum is mostly "green shifted" showing 
a maximum around $\gamma\kappa_B T$ with respect to the original "red" BBR 
spectrum with a maximum around $\kappa_B T$. The next step is to derive the 
total number of diffused photons in the EF; this number can be obtained as 
follows: 
\begin{equation}
\frac{dN_1^*}{dt_1^*d\epsilon_1^*d\Omega_1^*d\epsilon_o^*d\Omega_o^*}=\frac
{dn_o^*}{d\epsilon_o^*d\Omega_o^*}\frac{d\sigma_C}{d\epsilon_1^*d\Omega_1^*}c
\end{equation}
where $\frac{d\sigma_C}{d\epsilon_1^*d\Omega_1^*}$ is the Compton differential 
cross section 
\be
\frac{d\sigma_C}{d\epsilon_1^*d\Omega_1^*}=\frac{r_o^2}{2}\Big(
\frac{\epsilon_1^*}{\epsilon_o^*}\Big)^2\Big(\frac{\epsilon_o^*}{\epsilon_1^*}
+\frac{\epsilon_1^*}{\epsilon_o^*}-\sin^2\theta_{sc}^*\Big)\delta\Big(
\epsilon_1^*-\frac{\epsilon_o^*}{1+\epsilon_o^*(1-\cos\theta_{sc}^*)/mc^2}\Big
)
\ee
$c$ is the speed of light and the scattering angle must be expressed as a 
function of the other angles involved, i.e. the incoming $\theta_o^*
,\varphi_o^*$ and the outcoming $\theta_1^*,\varphi_1^*$ angles
\be
\cos\theta_{sc}^*=\sin\theta_o^*\sin\theta_1^*(\cos\varphi_o^*\cos\varphi_1^*+
\sin\varphi_o^*\sin\varphi_1^*)+\cos\theta_o^*\cos\theta_1^*~~~~~~~~~.
\ee 
The last step is to derive the final exact ICS differential number 
distribution in the Laboratory Frame (where the scattered photons will be 
observed) by using the inverse Lorentz transformations; the result is:
$$
\frac{dN_1}{dt_1 d\epsilon_1 d\Omega_1}=\frac{r_o^2 c(1-\beta\cos\theta_1)
\epsilon_1^2}{c^3 h^3}\int_{\Omega_o^*}\frac{1}{\exp\Big(\frac
{\gamma^2\epsilon_1(1-\beta\cos\theta_1)(1+\beta\cos\theta_o^*)}
{\kappa_B T[1-\gamma\epsilon_1(1-\beta\cos\theta_1)(1-\cos\theta_{sc}^*)/mc^2]}
\Big)-1}\cdot
$$
$$
\cdot\Bigg[\frac{1}{1-\gamma\epsilon_1(1-\beta\cos\theta_1)
(1-\cos\theta_{sc}^*)/mc^2}-\frac{\gamma\epsilon_1(1-\beta\cos\theta_1)
(1-\cos\theta_{sc}^*)}{mc^2}+
$$
\be
+\cos^2\theta_{sc}^*\Bigg]\frac{d\Omega_o^*}{\Big[1-\gamma\epsilon_1
(1-\beta\cos\theta_1)(1-\cos\theta_{sc}^*)/mc^2\Big]^2}
\ee
where from eq.4 $\cos\theta_{sc}$ must be expressed as a function of 
$\theta_o^*,\theta_1^*,\varphi_o^*,\varphi_1^*$. This "blue shifted" 
distribution is different from the previous ones because now the expected peak 
around a given $\epsilon_1$ energy has been spread into a wide plateau from 
$\kappa_B T$ up to $\gamma^2\kappa_B T$ energies. The most general ICS 
differential distribution can be obtained by means of a numerical integration 
of eq.5 over $\Omega_o^*$ but as we are more interested in high energy 
phenomena we also show how to simplificate the above formula in this cases. 
First of all we  remind that in particle accelerators we are dealing with 
ultrarelativistic particles, i.e. $\gamma\gg 1$; moreover for most 
astrophysical problems the photon energy in the EF is much smaller than the 
electron rest mass, i.e. $\epsilon_o^*\ll mc^2$; so it is often possible to 
approximate the Compton differential cross section by the Thomson one; as a 
third further approximation we may consider the ultrarelativistic-Thomson 
limit where the two previous conditions $\gamma\gg 1$ and $\epsilon_o^*\ll 
mc^2$ are both satisfied. Let us discuss these three different expansions. In 
the first case ($\gamma\gg 1$) the BBR photons, in the EF, are pratically all 
incident head-on so the incident angle $\theta_o^*$ can be approximately 
written as $\theta_o^*\approx\pi-\frac{1}{\gamma}$. Consequently the 
scattering angle is related to the $\theta_1^*$ angle by the simple formula 
$\theta_{sc}^*+\theta_1^*\simeq\pi$ and $\cos\theta_{sc}^*\simeq -
\cos\theta_1^*$ within $1/\gamma$. Moreover the kinematics of the ICS shows 
that the scattered radiation, in the LF, is strongly concentrated in a narrow 
cone $\theta_1\simeq 1/\gamma$ along the direction of motion of the 
ultrarelativistic particle. The ICS differential distribution in eq.5, in the 
ultrarelativistic expansion, reduces to an analytical expression 
$$
\frac{dN_1}{dt_1 d\epsilon_1 d\Omega_1}=\frac{2\pi\kappa_B T r_o^2 c}
{c^3 h^3\beta\gamma^2}\epsilon_1\ln\Bigg[\frac{1-\exp\Big( 
\frac{-2\gamma^2\epsilon_1(1-\beta\cos\theta_1)}{\kappa_B T [1-\epsilon_1(1+
\cos\theta_1)/2 mc^2\gamma]}\Big)}
{1-\exp\Big(\frac{-\epsilon_1(1-\beta\cos\theta_1)}{2\kappa_B T 
[1-\epsilon_1(1+\cos\theta_1)/2 mc^2\gamma]}\Big)}\Bigg]\cdot
$$
\be
\cdot\Bigg[1-\frac{\epsilon_1(1+\cos\theta_1)}{2 mc^2\gamma}\Bigg]^{-1}\Bigg[
\Big(1-\frac{\epsilon_1(1+\cos\theta_1)}{2 mc^2\gamma}\Big)^{-1}-
\frac{\epsilon_1(1+\cos\theta_1)}{2 mc^2\gamma}+
\Big(\frac{\cos\theta_1-\beta}{1-\beta\cos\theta_1}\Big)^2\Bigg]
\ee
where $1-\beta\simeq 1/2\gamma^2$ and $d\Omega_1\simeq 2\pi\theta_1d\theta_1$. 
The second possible approximation is the Thomson limit $\epsilon_o^*\ll mc^2$. 
In this case we can neglect all terms of order $\epsilon_o^*/mc^2$ with 
respect to 1 in eq.5 and the resulting differential distribution becomes
$$
\frac{dN_1}{dt_1 d\epsilon_1 d\Omega_1}=\frac{2\pi\kappa_B T r_o^2 c}{(ch)^3}
\epsilon_1\Bigg[\ln\Bigg[\frac{1-\exp\Big(-
\frac{\gamma^2\epsilon_1(1-\beta\cos\theta_1)(1+\beta)}{\kappa_B T}\Big)}
{1-\exp\Big(-\frac{\gamma^2\epsilon_1(1-\beta\cos\theta_1)(1-\beta)}
{\kappa_B T}\Big)}\Bigg]\cdot
$$
$$
\cdot\Bigg(1+\Big(\frac{\cos\theta_1-\beta}{1-\beta\cos\theta_1}\Big)^2\Bigg)
+\Bigg(1-3\Big(\frac{\cos\theta_1-\beta}{1-\beta\cos\theta_1}\Big)^2\Bigg)
\cdot
$$
\be
\cdot\int_{o}^{\pi}
\ln\Big[1-\exp\Big(-\frac{\gamma^2\epsilon_1}
{\kappa_B T}(1-\beta\cos\theta_1)(1+\beta\cos\theta_o^*)\Big)
\Big]\sin\theta_o^*\cos\theta_o^* d\theta_o^*\Bigg]
\ee
Indeed, from the previous eq.6-7, we can get the last expansion, the 
Thomson-ultrarelativistic formula. It stems from assuming $\gamma\gg1$ and 
$\epsilon_o^*\ll mc^2$ at the same time. This means that the ICS differential 
distribution can be obtained by setting $\epsilon_1\ll mc^2\gamma$ in eq.6 and 
by setting $\gamma\gg 1$ in eq.7. In the first derivation we obtain:
\begin{equation}
\frac{dN_1}{dt_1 d\epsilon_1 d\Omega_1}=\frac{2\pi\kappa_B T r_o^2 c}
{c^3 h^3\beta\gamma^2}\epsilon_1\ln\Bigg[\frac{1-\exp\Big(-\frac{2\gamma^2
\epsilon_1(1-\beta\cos\theta_1)}{\kappa_B T}\Big)}{1-\exp\Big(-\frac{\epsilon_1
(1-\beta\cos\theta_1)}{2\kappa_B T}\Big)}\Bigg]\Bigg[1+\Big(\frac{\cos
\theta_1-\beta}{1-\beta\cos\theta_1}\Big)^2\Bigg]
\end{equation}
In the second derivation the integral contained in eq.7 can be simplified 
remembering that $\theta_o^*\simeq\pi-\frac{1}{\gamma}$ so $\sin\theta_o^*
\cos\theta_o^*\simeq 1/\gamma$ and the integrand is significatively different 
from zero only inside a thin cone whose aperture is just $1/\gamma$. Thus the 
result is a formula differing from eq.8 only by a factor of order 
$\frac{1}{\gamma^2}$ and in our assumptions this term is completely 
negligible. So the still analytical eq.8 is our ultrarelativistic-Thomson 
distribution. In fig.1 we show the surprising metamorphosis of a 
Planckian Black Body into a taller (by a suppression factor $\gamma^{-2}$) 
serial of smooth truncated hills at larger relativistic regimes. This process 
is mainly due to the overlapping of a series of blueshifted spectra obtained 
at different angular directions; the softer photons are the more isotropic 
ones while the harder photons are strongly anisotropic and located in the 
inner cone of the beam. The "Rayleigh" regions of the differential "BBR" 
spectra 
overlap each other even if calculated at different $\theta_1$ angles 
($0<\theta_1<\pi$) while their peaks are higher and more and more blueshifted 
for $\theta_1$ angles approaching zero. Consequently the exponential decay of 
the ICS spectrum on the right side reflects the "Wien" regions behaviour of 
the angular "BBR" spectra at $\theta_1\approx0$. We give here a qualitative 
summary of the ICS energy spectrum behaviour in the 
whole $\epsilon_1$ energy range. For $\epsilon_1\ll \kappa_B T/\gamma^2$ the 
spectrum exhibits a linear growth (like the Rayleigh region of a BBR one); for 
energies $\kappa_B T/\gamma^2<\epsilon_1<\kappa_B T$ the linear behaviour is 
modified by a logaritmic correction and the growth is proportional to 
$\epsilon_1\ln(4\gamma^2\kappa_B T/\epsilon_1)$; for $\kappa_B T<\epsilon_1<
4\gamma^2\kappa_B T$ the spectrum is spread into a very slowly linearly 
decreasing plateau up to $\epsilon_1\approx 4\gamma^2\kappa_B T$; for 
$\epsilon_1>4\gamma^2\kappa_B T$ the spectrum 
decays as $\epsilon_1\exp(-\epsilon_1/4\gamma^2\kappa_B T)$ (we remember the 
reader that the Wien region of a BBR distribution decays as 
$\epsilon_1^2\exp(-\epsilon_1/\kappa_B T)$). We remind that the commonly and 
widely applied analytical formulae on ICS are the ones derived by F.Jones 
[3] and based on ICS onto monochromatic and isotropic radiation; this
ficticious and artificial "BBR" leads to a final spectrum in disagreement 
with the experimentally observed ones. In the next two sections we show 
that we can evaluate in the right way ICS spectra in high energy physics 
and astrophysics by means of our formulae.

\section*{The ICS spectrum in LEP experiments}

We could directly verify the validity of present results by a (successful) fit 
of experimental ICS spectra obtained at LEP by A.Melissinos group [7] and 
by G.Diambrini-Palazzi group [8] (at a higher degree of precision). Indeed the 
LEP vacuum pipe can be considered as a black body cavity at room temperature 
($T\approx 291~K$); hence the electromagnetic radiation in thermal equilibrium 
is scattered and beamed ahead by 
the electron (and positron) bunches whose energy is $E_e=~45.6~GeV$. At 
this energy corresponds a Lorentz factor $\gamma=8.92\cdot 10^4$ so we can 
apply our 
ultrarelativistic formulae for ICS to evaluate the interesting region of the 
spectrum. We compare our results with the Montecarlo simulations performed at 
LEP and able to fit the ICS effect in the experimental data. We compute either 
the 
Thomson and the Compton spectra in order to show the differences between the 
two curves and with respect to the Montecarlo simulation. The spectra can be 
obtained by a numerical integration over the $\theta_1$ angle in expressions 
6-8 respectively for Compton or Thomson limit:
\begin{equation}
\frac{dN_{1(TH,KN)}}{d\epsilon_1}=N_e\int_{\Omega_1}\frac{dN_{1(TH,KN)}}{dt_1 
d\epsilon_1 d\Omega_1} d\Omega_1\Delta\tau
\end{equation}  
where $N_e\simeq 1.37\cdot 10^{11}$ is the number of particles in a bunch and 
$\Delta\tau=\frac{l}{c}\simeq 2\cdot 10^{-6}~s$ is the flight time in the 
LEP straight section $l\simeq 600~m$ [9]. The three ICS spectra, derived from 
our equations and from the Montecarlo simulation are shown in fig.2 and 
labelled respectively as:\par
1) M: Montecarlo simulation (ref.[9] fig.6)\par
2) T: Thomson approximation, from eq.8\par
3) C: Compton approximation, from eq.6\par   

Note that our Compton spectrum shows a good agreement with the 
Montecarlo simulation while the Thomson one, at high energies, is a factor 3 
higher. This overestimate of the Thomson spectrum is clearly related to the 
independence of the Thomson cross section with respect to the photon energy.
We notice 
that our ICS spectrum, in the Compton limit, is nearly coincident with the 
Montecarlo approximation at low energy but for higher energies ($>GeV$) it is 
a 26$\%$ above. This discrepancy seems not to be related to our approximations 
because from our data we obtain, for the total event number, $N_1=2.646$ a 
value pratically coincident (within $0.1\%$) with $N_1=2.65$ found by Di 
Domenico [9] by Montecarlo simulations. The difference might be due to 
the Montecarlo method whose statistical procedure implies a smaller number of 
events at higher energies. However this discrepancy does not affect the beam 
lifetime because in its evaluation only the number rate $dN_1/dt_1$ is 
involved. We remind that the ICS may also be used to study the bunch internal 
structure and 
the result of a coherent emission by the charges, in this case the optimal 
experimental set up is reached when the relativistic bunches are hit by 
collinear back photon emitted by a laser [11].

\section*{The ICS in Astrophysics}

The $X-\gamma$ astronomy traces mostly the presence of relativistic electrons 
(or, at lower level, of relativistic nuclei) by their synchrotron radiation or 
their ICS onto infrared, interstellar or cosmic radiation. Moreover the same 
ICS may become the main slowing down process for relativistic charges once 
magnetic field energy densities $\rho_B$ are below the corresponding cosmic 
photon energy densities $\rho_{BBR}$. This situation generally occurs in 
extragalactic spaces where $\rho_B\ll\rho_{BBR}$. In this framework cosmic 
rays electrons around SN1006 have been recently discovered indirectly by their 
non 
thermal X-ray emission due to synchrotron radiation from ultrarelativistic 
electrons ($\gamma\geq 10^8$) [12]. It is therefore of great interest and 
actuality to 
provide not just an order of magnitude for the corresponding ICS $\gamma$ ray 
flux but also for its detailed spectrum for such relativistic and 
ultrarelativistic cosmic rays electrons. We show in fig.3 these ICS spectra 
for $\gamma=10^4,10^6,10^8$. It is important to note the existence of a high 
energy transition in the ICS from the Thomson to the Compton behaviour. This 
change 
occurs dramatically at the highest energy range of the spectra. The $\gamma$ 
Lorentz factor for Compton behaviour occurs for electrons ($e^-+
\gamma\rightarrow e^-+e^++e^-$) and for protons ($\gamma+p\rightarrow n+\pi^+, 
p+\pi^o$, ...) in ICS with $T=2.73~K$ BBR respectively at huge energy values
\be
\gamma_e=\frac{2m_e c^2}{<\epsilon_o>}=\frac{m_e c^2}{\kappa_B T}\approx 1.6
\cdot 10^9~~~;~~~ 
\gamma_p\approx\frac{2m_{\pi}c^2}{<\epsilon_o>}\simeq\frac{m_{\pi}c^2}
{\kappa_B T}=2.4\cdot 10^{11}
\ee
The ICS behaviour in the Compton limit may be qualitatively predicted keeping 
in mind that the ICS energy spectrum falls off at photon energies near 
$\epsilon_1\approx\gamma^2\kappa_B T$ and the energy conservation calls for a 
cut-off of the extension of the ICS spectrum plateau from $\kappa_B T$ to $m 
c^2\gamma$ instead of reaching the usual extreme value $\gamma^2\kappa_B T$. 
Let us better understand this ICS behaviour change from Thomson to Compton 
regime as follows: as long as the incident photon, in the EF, has a 
characteristic energy $\epsilon_o^*\approx\gamma\kappa_B T<m c^2$ the diffused 
photon, in the EF, mantains most of its original energy and it is spread 
around nearly isotropically. Therefore, once the spectrum is reviewed in the 
LF after the Lorentz boost, the previous different angular distribution of the 
photons in the EF becomes, in the LF, a different energy distribution of the 
same photons. The angular 
integral of these differential "BBR" spectra $\frac{dN_1}{dt_1 d\epsilon_1 
d\Omega_1}$, whose "Rayleigh" regions overlap, produces in the energy range 
$\kappa_B T<\epsilon_1<\gamma^2\kappa_B T$ the wider plateau shown in fig.4. 
However, at Compton regime, when $\gamma\kappa_B T>m c^2$ the differential 
cross section $\sigma\sim\sigma_T\cdot\frac{m}{\epsilon_o^* 
(1-\cos\theta_{sc}^*)}$ leads, in the EF, to a 
diffused and anisotropic photon distribution which becomes more and more 
beamed in a "Compton cone" at angles $\theta_C^*\leq\sqrt{\frac{m 
c^2}{\kappa_B T}}\sqrt{\frac{1}{\gamma}}$ (not to be confused with the thinner 
"Lorentz" angles $\theta_L^*\leq\frac{1}{\gamma}$ related to the boost 
transformations). All the photons diffused outside the $\theta_C^*$ cone can 
reach in the LF final energies of order $\epsilon_1\approx\gamma^2\kappa_B T
(1-\beta\cos\theta_o^*)(1-\beta\cos\theta_{sc}^*)\simeq 2\gamma^2\kappa_B T
(1-\beta\cos\theta_{sc}^*)\leq mc^2\gamma$. On the other side, around and 
inside the 
Compton cone $\theta_C^*$ but far from the edge of the inner Lorentz cone 
$\theta_L^*$ (for example we may consider $\frac{1}{2}\sqrt{\frac{m 
c^2}{\gamma\kappa_B T}}<\theta_{sc}^*<\sqrt{\frac{m c^2}{\gamma\kappa_B T}}$ 
where the majority ($>75\%$) of the scattered photons is found) the final 
avalaible energy, in the LF, will be $\epsilon_1\sim\gamma^2\kappa_B T(1-
\beta\cos\theta_{sc}^*)\geq\kappa_B T+\frac{m c^2}{8}\gamma$. In conclusion 
the Compton behaviour of the ICS spectra, derived analitically from the exact 
formula 5 and described in fig.5, piles up the photons near the edge 
energies in the range $\epsilon_1\leq m c^2\gamma$ and leads 
to an unexpected peak higher and higher at those highest energy values as the 
Lorentz factor $\gamma\gg\frac{m c^2}{\kappa_B T}$ increases. It is also 
important 
to notice that this extreme Compton regime may arise in a different way: for 
example in non relativistic or relativistic cases when $\kappa_B T\gg m c^2$. 

\section*{The main scenarios for ICS applications}

For a synthetic but complete picture of all the above described ICS behaviours 
we suggest to define the following characteristic regimes each labelled by the 
relevant parameters involved.\newline
1) ($\gamma\sim 1$, $\kappa_B T\ll\frac{m c^2}{\gamma}$) {\bf The
non-relativistic Compton scattering limit onto cold-warm BBR} (fig.6). 
Here the simplest ICS spectrum becomes a self similar Planck spectrum; its 
reflectivity efficiency depends on the usual quantities $n_{\gamma}$, 
$\sigma_T$, ... One of the mostly celebrated application is the 
Sunyaev-Zeldovich effect in cosmology.\newline
2) ($\gamma\gg 1$, $\kappa_B T<\frac{m c^2}{\gamma}$) {\bf The ICS at 
relativistic Thomson limit onto cold-warm BBR} (fig.1-2). This is 
the case with most application in high energy physics and astrophysics (LEP I, 
LEP II, GRBs, cosmic rays at energies $E_e\leq 10^{14}~eV$ and $E_p\leq 
10^{17}~eV$,...). Now the ICS spectrum deviates from a pure Planckian spectrum 
leading to a "cut-hill" spectrum with its smooth edge covering the energies 
from $\epsilon_1\sim\kappa_B T$ up to photon energies 
$\epsilon_1\sim\gamma^2\kappa_B T$. This "simple" ICS spectrum 
may be ruling the spectra of ICS by charge (electron) jets ($\gamma\sim 10^3-
10^4$) onto nearby stellar companion photons ($\epsilon_o\sim 0.5~eV$) leading 
to successful gamma jets able to explain the integral spectra of GRBs [13]. 
This arguments have widely been developed by the authors 
and are still under consideration [14]. Because of the "smooth" behaviour of 
the edge of the spectrum there is not a one to one relation between the cosmic 
rays electron or proton spectral index and the final ICS $\gamma$ rays 
spectral index. Another scenario where ICS leads to $\gamma$ rays takes 
place in extragalactic blazars which eject beamed cosmic rays (protons, 
nuclei,...) for huge distances. Their ICS onto 2.73 K BBR may also lead to 
collinear $\gamma$ jets; their energy is dominated by the electron presence 
(over the nuclei one) in the cosmic rays jets. Such gamma jet is the large 
scale version of the minijet model plus ICS considered by the authors for 
GRB production in galactic binary systems.\newline
3) ($\gamma\gg 1$, $\frac{m c^2}{\gamma}\leq\kappa_B T\ll m c^2$) 
{\bf The ICS at relativistic Compton limit onto cold-warm BBR} (fig.7). These 
energy windows are relevant for the highest cosmic rays interactions 
onto BBR ($E_e\gg 10^{14}~eV$, $E_p\geq 10^{20}~eV$). The ICS spectrum 
exhibits a pile up of photons at the edge of the highest $\epsilon_1$ energies 
leading to a peaked maximum at $\epsilon_1\sim m c^2\gamma$ and a sharp cut 
off at energies just above it. The presence of such a peak has important 
consequences in the cosmic rays-$\gamma$ rays links. The primordial incident 
cosmic rays spectrum at highest energies (power law, ...) leaves its original 
inprint into a similar "photocopy" $\gamma$ ray spectrum. Moreover the new 
born high energies $\gamma$ rays ($\epsilon_1> 100~TeV$) may also successfully 
interact with the BBR photons (electron pairs production) leading to  
electron-proton cascades or electromagnetic showers at lower and lower 
energies. The showers will arrest their growth as soon as the last degraded 
photon energies will reach the threshold $E_c=\frac{2m c^2}{\kappa_B T} m c^2=
10^{15}~eV$. It is important to remind that the $\gamma-\gamma\rightarrow e^++
e^-$ process at ultrarelativistic limit has a very similar cross section (
Breit-Wheeler, 1934) as the $e^+e^-$ annihilation (Dirac, 1930) or the Klein-
Nishina (1929) cross section. One of the consequences is the expected 
cosmic background presence of 
relic TeV $\gamma$ rays components (due to the direct cosmological ICS or 
due to its secondary showers at energies $\epsilon_1\leq 100~TeV$). This 
argument will be discussed in detail elsewhere.\newline
4) ($\gamma\sim 1$, $\kappa_B T\geq m c^2$) {\bf The non relativistic 
Compton scattering onto "hot" BBR} (fig.8). The resulting ICS spectrum 
deviates from the 
original one and exhibits a small peak at energies $\epsilon_1\sim\frac{m 
c^2}{2}$. The effect as above is related to the energy dependence of the 
Klein-Nishina cross section.\newline 
5) ($\gamma\gg 1$, $m c^2\ll\kappa_B T\ll m c^2\gamma$) {\bf The 
relativistic ICS onto "hot" BBR}. The spectrum shows a new "plateau" 
extending from 
$\epsilon_1\sim\kappa_B T$ energies up to $\epsilon_1\sim m c^2\gamma$ 
energies with a marked peak at $\epsilon_1\sim m c^2\gamma$ energies and a 
sharp cut off above, this behaviour is similar of point 3).\newline
6) ($\gamma\gg 1$, $\kappa_B T=m c^2\gamma$) {\bf The equilibrium 
regime of ICS onto "hot" BBR} (fig.9). The spectrum shows a marked peak at 
$\epsilon_1\sim m c^2\gamma$ energies with a sharp cut-off above. The known 
astrophysical scenarios where such ICS may play a role could be the earliest 
hottest (thermal equilibrium) cosmological epochs ($t\leq 1~s$, $T>~MeV$) and 
in the hot thermal cores of supernovae explosions ($\kappa_B T\geq 5~MeV$). It 
is interesting to notice the nature of the non equilibrium ICS spectrum and 
it may be 
worthfull to derive the exact kinetic equations (due to such ICS) for the 
multicomponent thermal bath of the early universe as well as the SN 
explosive 
processes. Finally if the fireball model is a real event as the one needed to 
explain the GRB puzzle then such ICS (onto the last layers of the fireball 
explosion) would be smeared out by multiscattering during last stages of 
the fireball into a 
final nearby thermal GRB spectrum. We do not recognize such a presence of 
thermal inprint in GRB [13].\newline
7) ($m c^2\ll m c^2\gamma\ll\kappa_B T$) {\bf The ultrarelativistic 
"ultrahot" ICS} (fig.10). This ICS spectrum exhibits a peaked spectrum as 
usual at $\epsilon_1\sim m c^2\gamma$ energies but also a significative 
decaying component at higher energies. We can see a "shoulder" which may 
become, in the 
extreme cases $\kappa_B T\gg m c^2\gamma$, a slowly decaying "plateau". Such a 
peculiar process (at the present) seems very hypothetical but we have shown 
it in order to cover the entire range of possible ICS behaviours. Possible 
exotic 
scenarios where such extreme conditions (5,6,7) may take place efficiently 
occur also near miniblackhole evaporations in the early Universe where the 
cosmological BBR has a temperature lower or comparable or even greater than 
the corresponding earliest miniblackhole temperature ($\kappa_B T_{mbh}
\approx m c^2\gamma$). This hybrid thermal bath, depending on miniblackhole 
primordial masses and distribution, may be dominant in non standard early 
cosmology bariogenesis and even in later cosmological nucleosynthesis.

\section*{Conclusions and Applications}

The developed ICS formulae have a wide range of applications; in particular in 
fitting LEP I, LEP II experimental data and in understanding recent GRB 
puzzling spectra (in this last case we modified eq.8 in order to take into 
account the diluted and anisotropic BBR spectrum seen from the jet and we 
assumed a ring-like photon source ([14] eq.20). Moreover the extreme spectra 
of ICS at relativistic Compton 
limit onto cold-warm BBR, its peak at highest energies, may be probed at LEP 
I, LEP II using a diffused thermal optical light (a flash) in the beam pipe 
during the bunch crossing. Finally the recent evidence for cosmic rays 
electron at energies above hundreds TeV by their observed synchrotron 
radiation at soft X spectra implies the coexistence of a low but nearly 
detectable component of high $\gamma$ cosmic rays at energies $E_{\gamma}\leq 
m c^2\gamma\approx 100~TeV$. Because of arguments in 3) the cosmic rays 
electrons spectrum ($dN/dE_e\sim E_e^{-2}$) will be reflected also in the 
"photocopy" $\gamma$ spectrum. Their total energy flux $\phi_{\gamma}$ will be 
of the same order of magnitude of the X rays one:
\be
\phi_{\gamma}\leq\phi_X\frac{\rho_{BBR}}{\rho_B}\leq\frac{1}{2}\phi_X\Big(
\frac{B}{6\mu G}\Big)^{-2} 
\ee
The corresponding $\gamma$ ray flux number will be extremely 
suppressed because of the $X-\gamma$ energy ratio
\be
\frac{dN_{\gamma}}{dN_X}\simeq\frac{\phi_{\gamma}}{\phi_X}\frac{<E_X>}
{<E_{\gamma}>}\sim 2\cdot 10^{-11}\Big(\frac{B}{6\mu G}\Big)^{-1}\Big(
\frac{E_e}{100~TeV}\Big)
\ee
and it follows that the needed area for such a low flux is possibly below the 
present air shower arrays sensibility. However an accurate direction-flux 
correlation might be able to observe in a near future a tiny 100 TeV flux. We 
suggest to seriously consider a detailed observational program in order to 
detect above the tiny 100 TeV $\gamma$ flux from SN1006 and possibly the other 
suggested candidates (Cas A, IC443, Tycho SN, ...) from SNRs. Their presence 
at the above fluxes is a necessary and compelling consequence of fundamental 
QED and cosmological BBR theories combined in present ICS models. Finally we 
interpret 
the two X lobes around SN1006 as generated by an electron jet scattering onto 
the relic giant shell contrary to the more popular idea of a Fermi shock 
acceleration mechanism. In these beaming models one should also expect a rare  
and strongly time dependent TeV $\gamma$ rays "burst" which could be 
observed in a short period of time (hours) once the observer is inside the 
thin jet cone direction, with an amplified integral intensity by a factor 
$\gamma$ Lorentz larger then a diffused spherical source. Well known 
candidates are blazars or quasars as 3C279 , AGNs, as NGC 3079, or most recent 
TeV Mrk sources. Similar arguments leaded us to expect a low variable relic 
background of gamma (ten TeV) noise due to the pile up of cosmic integral ICS 
gamma rays and their  electromagnetic cascades just below the electron pairs 
creation threshold on cosmic BBR.     

\section*{Acknowledgements}

We wish to thank Prof. G.Diambrini-Palazzi and Dott. A.Di Domenico for useful 
discussions and support. 

\section*{References}

[1]~Ginzburg V.L., Syrovatskii S.I.,~~~~~{\it Soviet Phys.JETP}~~{\bf 19},
~1255~(1964)\newline
~~[2]~Feenberg E., Primakoff H.,~~~~~{\it Phys.Rev.}~~{\bf 73},~449~(1948)
\newline
~~[3]~Jones F.C.,~~~~~{\it Phys.Rev.}~~{\bf 167},~1159~(1968) \newline
~~[4]~Blumenthal G.R., Gould R.J.,~~~~~{\it Rev.Mod.Phys.}~~{\bf 42},~237~
(1970) \newline
~~[5]~Longair M.S.,~~~"High Energy Astrophysics"~~Cambridge University \newline
Press~1981 \newline
~~[6]~Berezinskii V.S., Bulanov S.V., Dogiel V.A., Ginzburg V.L., \newline 
Ptuskin V.S.~~~"Astrophysics of Cosmic Rays"~~North Holland~1990 \newline
~~[7]~Dehning B., Melissinos A.C., Ferrone P., Rizzo C., von Holtey G., \newline
{\it Phys.Lett.}~~{\bf B249},~145~(1990) \newline
~~[8]~Bini C., De Zorzi G., Diambrini-Palazzi G., Di Cosimo G., \newline 
Di Domenico A., Gauzzi P., Zanello D.,~~~~~{\it Phys.Lett.}~~{\bf B262},~135~
(1991) \newline
~~[9]~Di Domenico A.,~~~~~{\it Particle Accelerators}~~{\bf 39},~137~(1992) 
\newline
~~[10]~Fargion D.,~~~in "The Dark Side of the Universe"~~~World~Scientific~Pub.
~pag.88-99~(1994) \newline
~~[11]~Fargion D., Konoplich R.V., Salis A.,~~~Preprint INFN 1115, 23/10/95 
Rome\newline 
[12]~K.Koyama et al., ~~~~{\it Nature}~~{\bf 378},~255~(1995)\newline
[13]~Fargion D., Salis A.,~~~~{\it Nucl.Phys.B~(Proc.Suppl.)}~{\bf 43},~269~
(1995)\newline
[14]~Fargion D.,Salis A.~~~~{\it Astrophysics and Space Science}~{\bf 231}:~
191-194~(1995)\newline
Fargion D., Salis A.,~~~~NATO Proc.C461, Kluwer Pub. 1995

\newpage

\centerline{\bf{Figure Captions}}

\begin{itemize}

\item{\bf Fig.1}: The ICS spectrum from eq.8 for T=291 K and non relativistic 
(BBR) $\gamma=1$ (dot), and relativistic $\gamma=10$ (dot dash), $\gamma=10^2$ 
(dash), $\gamma=10^3$ (continuous) Lorentz factor 

\item{\bf Fig.2}: The ICS spectra at LEP: Montecarlo simulation M (ref.9 
fig.6), Thomson approximation T (eq.8), Compton approximation C (eq.6)

\item{\bf Fig.3}: The ICS spectrum for T=2.73 K and $\gamma=10^4$ (dot dash), 
$\gamma=10^6$ (dot), $\gamma=10^8$ (continuous) Lorentz factor

\item{\bf Fig.4}: The ICS "Thomson" spectrum for T=2.73 K and $\gamma=10^4$ 
Lorentz factor

\item{\bf Fig.5}: The ICS spectrum evolution from "hill-like" Thomson spectra 
$\frac{\kappa_B T}{m c^2}=10^{-2}$ (dash), $10^{-1}$ (lower continuous), 1 
(dot dash) to Compton peaked spectra $\frac{\kappa_B T}{m c^2}=10$ (dot), 
$10^2$ (higher continuous) for $\gamma=10$ Lorentz factor

\item{\bf Fig.6}: The ICS Thomson spectra for T=291 K and $\gamma=1$ (BBR) 
(dot), $\gamma=3$ (dash) Lorentz factor

\item{\bf Fig.7}: The ICS Compton spectra for $\kappa_B T=2.3\cdot 10^{-2}$ and 
$\gamma=10^9$ Lorentz factor

\item{\bf Fig.18}: The ICS Compton spectrum for $\kappa_B T=m c^2$ and 
$\gamma=5$ Lorentz factor

\item{\bf Fig.9}: The ICS Compton spectrum for $\kappa_B T=5.11\cdot 10^6~eV$ 
and $\gamma=10$ Lorentz factor ($\kappa_B T=m c^2\gamma$)

\item{\bf Fig.10}: The Compton ICS spectrum for $\kappa_B T=10^3~m c^2$ 
(continuous), $\kappa_B T=10^4~m c^2$ (dot) and $\gamma=10$ Lorentz factor

\end{itemize}

\end{document}